\documentclass[epj]{svjour}

\usepackage{hyperref}
\usepackage{graphicx}
\usepackage{epsfig,epsf,bm}

\begin{document}
\title{Spin-gap phase in the extended $t-J$ chain}
\author{Yu-Wen Lee\inst{1}\and Yu-Li Lee\inst{2}}
\institute{Physics Department, Tunghai University, Taichung,
Taiwan, R.O.C., \email{ywlee@mail.thu.edu.tw} \and Physics
Department, National Changhua University of Education, Changhua,
Taiwan, R.O.C., \email{yllee@cc.ncue.edu.tw}}

\abstract{We study the one-dimensional deformed $t-t^{\prime}-J$
model in terms of the continuum field theories. We found that at
low doping concentration and far away from the phase separation
regime, there are two phases: the Luttinger liquid and the
Luther-Emery liquid, depending on $t^{\prime}/t<(t^{\prime}/t)_c$
or $t^{\prime}/t>(t^{\prime}/t)_c$, where $(t^{\prime}/t)_c>0$.
Moreover, the singlet superconducting correlations are dominant in
the Luther-Emery liquid.}

 \maketitle

\section{Introduction}

Strong electronic correlations are widely believed to be crucial
for the understanding of the anomalous properties of
high-temperature superconductors. A popular approach in this
context is the use of the $t-J$ model, with holes moving in an
antiferromagnetic (AF) spin background. (Here $t$ and $J$ are the
nearest-neighbor hopping amplitude and the AF exchange
interaction, respectively.) In recent years, the measurements of
angle resolved photoemission spectroscopy (ARPES) in
Sr$_2$CuO$_2$Cl$_2$\cite{WSM} indicate that it is necessary to
include a next-nearest-neighbor hopping term to explain the ARPES
data, which leads to the ``extended" $t-J$ model. Subsequent
efforts have concentrated on the effects of the extra hopping
terms on various properties of planar and ladder systems, such as
stripe stability\cite{TGT}, competition between stripes and
pairing\cite{WS}, spin-charge separation in two
dimensions\cite{MED}, and spin gap evolution in two-leg
ladders\cite{PCR}. Currently, it is well-established that a
positive value of the next-nearest-neighbor hopping $t^{\prime}$
enhances hole pairing, while the opposite occurs for $t^{\prime}$
negative\cite{WS}.

In the present paper, we study the deformed version of the
one-dimensional ($1$D) $t-t^{\prime}-J$ model (see the following).
Our main results are as follows: (i) Far away from the phase
separation regime ($J/t\ll 1$), there are two phases at low doping
concentration: the Luttinger liquid (LL) for
$t^{\prime}/t<(t^{\prime}/t)_c$ and the Luther-Emery (LE) liquid
(spin-gap phase) for $t^{\prime}/t>(t^{\prime}/t)_c$, where
$(t^{\prime}/t)_c>0$. (The LE liquid is a spin-gapped state with
one gapless charge mode.) (ii) The value of $(t^{\prime}/t)_c$ is
expected to depend on the hole concentration $x$ and the ratio
$J/t$. At low hole concentration ($x\ll 1$), however, it is
insensitive to the value of $J/t$ and is an increasing function of
$x$. (iii) The spin gap is opened in the LE liquid and thus both
$2k_F$ charge density wave (CDW) and singlet superconducting (SS)
correlations are enhanced. But it is the SS one which is dominant
in the LE liquid.

The rest of this paper is organized as follows: The deformed
$t-t^{\prime}-J$ model and the corresponding continuum theory are
introduced in Section \ref{ttj}. Section \ref{pd} is devoted to
the phase diagram of the $1$D $t-t^{\prime}-J$ model. We study the
properties of the ground state in the spin-gap phase in Section
\ref{sgp}. The last section is the conclusions and discussions of
our results.

\section{Deformed $t-t^{\prime}-J$ model}
\label{ttj}

We start with the $1$D $t-t^{\prime}-J$ model which is described
by the Hamiltonian
\begin{eqnarray*}
 H=H_h+H_J \ ,
\end{eqnarray*}
where
\begin{eqnarray}
 {\rm H}_h &=& -t\sum_j\left(\bar{c}_{j+1\alpha}^{\dagger}
           \bar{c}_{j\alpha}+{\rm H.c.}\right) \nonumber \\
           & & -t^{\prime}\sum_j\left(\bar{c}_{j+2\alpha}^{\dagger}
           \bar{c}_{j\alpha}+{\rm H.c.}\right) \ , \label{ht} \\
 {\rm H}_J &=& J\sum_j\left(\bm{S}_j\cdot \bm{S}_{j+1}-\frac{1}{4}~
           n_jn_{j+1}\right) \ . \label{hj}
\end{eqnarray}
Here the Hubbard operator\cite{Hub} $\bar{c}_{j\alpha}$ is given
by
\begin{equation}
 \bar{c}_{j\alpha}^{\dagger}=c_{j\alpha}^{\dagger}(1-n_{j-\alpha})
        \ , \label{hop1}
\end{equation}
where $c_{j\alpha}$ is the annihilation operator of electrons on
the site $j$ with spin $\alpha$, and $\alpha =\uparrow
,\downarrow$ correspond to spin up and down, respectively.
Moreover, $n_j=c^{\dagger}_{j\alpha}c_{j\alpha}$ and
$\bm{S}_j=\frac{1}{2}~c^{\dagger}_{j\alpha}(\bm{\sigma})_{\alpha
\beta}c_{j\beta}$. In the following, we shall take $t,J>0$. The
extra factor in the Hubbard operator is to impose the
no-double-occupancy condition which results from the strong
Coulomb repulsion between two electrons with opposite spins on the
same site. Inserting equation (\ref{hop1}) into $H_h$ leads to
four- and six-fermion interactions with the same strengths as the
hopping amplitudes, which defies the perturbative approach.

To overcome this difficulty, Chen and Wu proposed to replace the
Hubbard operator (\ref{hop1}) by the deformed one\cite{CW}
\begin{equation}
 \bar{c}_{j\alpha}^{\dagger}=c_{j\alpha}^{\dagger}(1-\Delta n_{j-\alpha})
        \ , \label{hop2}
\end{equation}
with the deformation parameter $0<\Delta \leq 1$. That is, a
non-zero probability to leak into the states with double occupancy
is now allowed. The deformed model has the advantage that as
$0<\Delta \ll 1$ and $J/t\ll 1$, all four- and six-fermion
interactions can be treated as perturbations and thus a
field-theoretical approach can be performed. It is hoped that both
the $t-t^{\prime}-J$ model ($\Delta =1$) and the deformed one
($0<\Delta \ll 1$) fall into the same universality class.

Now we would like to study the low-energy physics of the $1$D
deformed $t-t^{\prime}-J$ model away from the phase separation
regime. This can be achieved by studying the corresponding
continuum field theory. Inserting equation (\ref{hop2}) into
equations (\ref{ht}) and (\ref{hj}), expanding the electron
operator around the Fermi points by
\begin{equation}
 c_{j\alpha}\approx \sqrt{a_0}\left[e^{-ik_Fx}\psi_{L\alpha}(x)+
            e^{ik_Fx}\psi_{R\alpha}(x)\right] \ , \label{eop1}
\end{equation}
where $x=ja_0$ with $a_0$ being the lattice spacing and $k_F$ is
the Fermi momentum determined by the electron density $n$ via
$k_Fa_0=\frac{\pi}{2}n$, ($n=1$ corresponding to the
half-filling.) and then taking the continuum limit
($a_0\rightarrow 0$) by keeping $ta_0$, $t^{\prime}a_0$, and
$Ja_0$ finite, we found that the Hamiltonian describing the
dynamics of $\psi$-fermions, up to the four-fermion interactions,
is given by
\begin{eqnarray*}
 H_{\psi}=\int dx \ (H_c+H_s) \ ,
\end{eqnarray*}
where
\begin{eqnarray}
 H_c &=& \frac{\pi}{2}~v_{c0}:\left(J_LJ_L+J_RJ_R\right):+\lambda_c
     J_LJ_R \ , \label{hc1} \\
 H_s &=& \frac{2\pi}{3}~v_s:\left(\bm{J}_L\cdot \bm{J}_L+\bm{J}_R
     \cdot \bm{J}_R\right):+\lambda_s\bm{J}_L\cdot \bm{J}_R \ ,
     \label{hs1}
\end{eqnarray}
with
\begin{eqnarray}
 J_{L(R)} &=& :\psi^{\dagger}_{L(R)\alpha}\psi_{L(R)\alpha}: \ ,
          \nonumber \\
 \bm{J}_{L(R)} &=& \frac{1}{2}:\psi^{\dagger}_{L(R)\alpha}
          (\bm{\sigma})_{\alpha \beta}\psi_{L(R)\beta}: \ .
          \label{current1}
\end{eqnarray}
In the above, $:\cdots :$ stands for normal ordering with respect
to the Fermi points. The parameters appearing in equations
(\ref{hc1}) and (\ref{hs1}) are defined by $v_{c0} \! = \! v_F[ \!
1+2g_c/ \! (\pi v_F) \! ]$, $v_{s}=v_F[1+3g_s/(2\pi v_F)]$, and
\begin{eqnarray}
 & & \lambda_c=4\left\{\Delta \left(1-\Delta \frac{n}{2}\right)
           \left[t\cos{\left(\frac{\pi n}{2}\right)}+2t^{\prime}
           \cos{(\pi n)}\right]\right. \nonumber \\
 & & \left.+\frac{\Delta^2}{\pi}\sin{\left(\frac{\pi n}{2}\right)}
     \left[t\cos{(\pi n)}+4t^{\prime}\cos{\left(\frac{\pi n}{2}
     \right)}\cos{(2\pi n)}\right]\right. \nonumber \\
 & & \left.-\frac{J}{4}\cos^2{\left(\frac{\pi n}{2}\right)}\right\}
     \ , \label{cp1} \\
 & & \lambda_s=-16\left\{\Delta \left(1-\Delta \frac{n}{2}\right)
     \left[t\cos{\left(\frac{\pi n}{2}\right)}+2t^{\prime}\cos{(\pi n)}
     \right]\right. \nonumber \\
 & & \left.+\frac{\Delta^2}{\pi}~\sin{\left(\frac{\pi n}{2}\right)}
     \left[t+4t^{\prime}\cos{\left(\frac{\pi n}{2}\right)}\right]
     \right. \nonumber \\
 & & \left.-\frac{J}{4}\cos^2{\left(\frac{\pi n}{2}\right)}\right\}
     \ , \label{sp1}
\end{eqnarray}
where $v_F=2t\left(1-\Delta \frac{n}{2}\right)^2 \! \!
\sin{k_F}\times \! \!
\left[1+4\frac{t^{\prime}}{t}\cos{k_F}\right]$ is the Fermi
velocity, and
\begin{eqnarray*}
 & & g_c=2\Delta \left(1-\Delta \frac{n}{2}\right)\left[t
     \cos{\left(\frac{\pi n}{2}\right)}+2t^{\prime}\cos{(\pi n)}
     \right] \\
 & & +\left[t\cos{(\pi n)}+4t^{\prime}\cos{\left(\frac{\pi n}{2}
     \right)}\cos{(2\pi n)}\right] \\
 & & \times \frac{\Delta^2}{\pi}~\sin{\left(\frac{\pi n}{2}\right)}
     -\frac{J}{4} \ , \\
 & & g_s=-\frac{4}{3}\left\{2\Delta \left(1-\Delta \frac{n}{2}\right)
     \left[t\cos{\left(\frac{\pi n}{2}\right)}+2t^{\prime}\cos{(\pi n)}
     \right]\right. \\
 & & \left.+\left[t(2+\cos{(\pi n)})+4t^{\prime}\cos{\left(
     \frac{\pi n}{2}\right)}(2+\cos{(2\pi n)})\right]\right. \\
 & & \left.\times \frac{\Delta^2}{\pi}~\sin{\left(\frac{\pi n}{2}
     \right)}-\frac{3}{4}J\right\} \ .
\end{eqnarray*}
We have set $a_0=1$ in the above expressions. In equations
(\ref{cp1}) and (\ref{sp1}), the terms proportional to
$\Delta^2\sin{(\pi n/2)}$ arise from the six-fermion interactions
in $H_h$, whereas the four-fermion interactions in $H_h$
contribute to those proportional to $\Delta (1-\Delta n/2)$. (For
the details of derivation, see Appendix \ref{ham}.) Note that the
above equations are derived by assuming $n\neq 1$ and thus the
Umklapp processes are irrelevant operators which can be neglected.
The higher order interactions between $\psi$-fermions are
neglected in $H_{\psi}$. This is because in the weak-coupling
regime ($\Delta \ll 1$) they are irrelevant operators in the sense
of renormalization group.

Before exploring the physics contained by equations (\ref{hc1}),
(\ref{hs1}), (\ref{cp1}), and (\ref{sp1}), three points have to be
mentioned. First, the Umklapp processes are neglected, which
changes the low-energy physics of the charge sector at
half-filling drastically. Therefore, the following analysis about
the charge sector cannot be extrapolated to that case. Next, the
inclusion of $t^{\prime}$ modifies the dispersion relation of free
electrons. It is possible that there are four Fermi points instead
of two upon hole or electron doping. The low-energy physics in
this case will be similar to that of the two-band system. However,
this will occur only when $|t^{\prime}/t|>0.25$ and the doping
concentration exceeds some critical value. On the other hand, the
above derivation starts with the one-band assumption --- equation
(\ref{eop1}). Thus, we shall apply equations (\ref{hc1}),
(\ref{hs1}), (\ref{cp1}), and (\ref{sp1}) to the following
situations: (i) finite doping concentrations for
$|t^{\prime}/t|<0.25$ and (ii) light doping for
$|t^{\prime}/t|>0.25$. Finally, equations (\ref{cp1}) and
(\ref{sp1}) are valid only when $\Delta \ll 1$ and $J/t\ll 1$.
Since the $1$D $t-J$ model is phase separated as $J/t\geq
O(1)$\cite{EKL}, the Hamiltonian $H_{\psi}$ describes the
low-energy physics in the region far away from the place where the
phase separation occurs.

\section{Phase diagram}
\label{pd}

Because of the spin-charge separation, the dynamics of the charge
and spin sectors can be discussed separately. We first consider
the charge sector. In terms of the bosonization formulas,
\begin{eqnarray}
 J_L+J_R &=& \sqrt{\frac{2}{\pi}}~\partial_x\Phi_c \ , \nonumber
         \\
 J_L-J_R &=& \sqrt{\frac{2}{\pi}}~\partial_x\Theta_c \ ,
         \label{bos1}
\end{eqnarray}
the Hamiltonian $H_c$ can be written as
\begin{equation}
 H_c=\frac{v_c}{2}:\left[K_c(\partial_x\Theta_c)^2+\frac{1}{K_c}
    (\partial_x\Phi_c)^2\right]: \ , \label{hc2}
\end{equation}
where $v_c=v_{c0}\sqrt{1-[\lambda_c/(\pi v_{c0})]^2}$ and
\begin{equation}
 K_c=\sqrt{\frac{1-\lambda_c/(\pi v_{c0})}{1+\lambda_c/(\pi v_{c0})}}
    \ . \label{cp2}
\end{equation}
Accordingly, the charge sector is gapless away from half-filling.

Next we turn into the spin sector. The relevancy of the coupling
$\lambda_s$ in $H_s$ can be determined by the one-loop RG
equation, which is given by
\begin{equation}
 \frac{d\lambda_s(l)}{dl}=\frac{1}{2\pi}\lambda_s^2(l) \ .
      \label{rg1}
\end{equation}
For simplicity, we have set $v_s=1$ in equation (\ref{rg1}). From
equation (\ref{rg1}), we see that for $\lambda_s(0)<0$,
$\lambda_s$ flows to zero and the spin excitations are gapless. On
the other hand, $\lambda_s(l\rightarrow \infty)\rightarrow \infty$
as $\lambda_s(0)>0$. In the latter case, the low energy physics
can be elucidated by abelian bosonization. Using the bosonization
formulas
\begin{eqnarray}
 J_L^+J_R^-+J_L^-J_R^+ &=& -\frac{1}{2\pi^2a_0^2}\cos{\left(
             \sqrt{8\pi}\Phi_s\right)} \ , \nonumber \\
 J_L^z+J_R^z &=& \frac{1}{\sqrt{2\pi}} \ \partial_x\Phi_s \ ,
             \nonumber \\
 J_L^z-J_R^z &=& \frac{1}{\sqrt{2\pi}} \ \partial_x\Theta_s \ ,
             \label{bos2}
\end{eqnarray}
where $J_{L(R)}^{\pm}=J_{L(R)}^x\pm iJ_{L(R)}^y$, the $\lambda_s$
term becomes
\begin{eqnarray*}
 \lambda_s\bm{J}_L\cdot \bm{J}_R &=& \lambda_s\left\{\frac{1}
          {8\pi}\left[(\partial_x\Phi_s)^2-(\partial_x\Theta_s)^2
          \right]\right. \\
          & & \left.-\frac{1}{4\pi^2a_0^2}\cos{\left(\sqrt{8\pi}
          \Phi_s\right)}\right\} \ .
\end{eqnarray*}
When $\lambda_s$ is positive, the cosine term becomes relevant and
$\langle \Phi_s\rangle =\sqrt{\pi /2}~l$ where $l$ is some
integer. This results in a spin gap.
\begin{figure}
\begin{center}
\includegraphics[width=0.9\columnwidth]{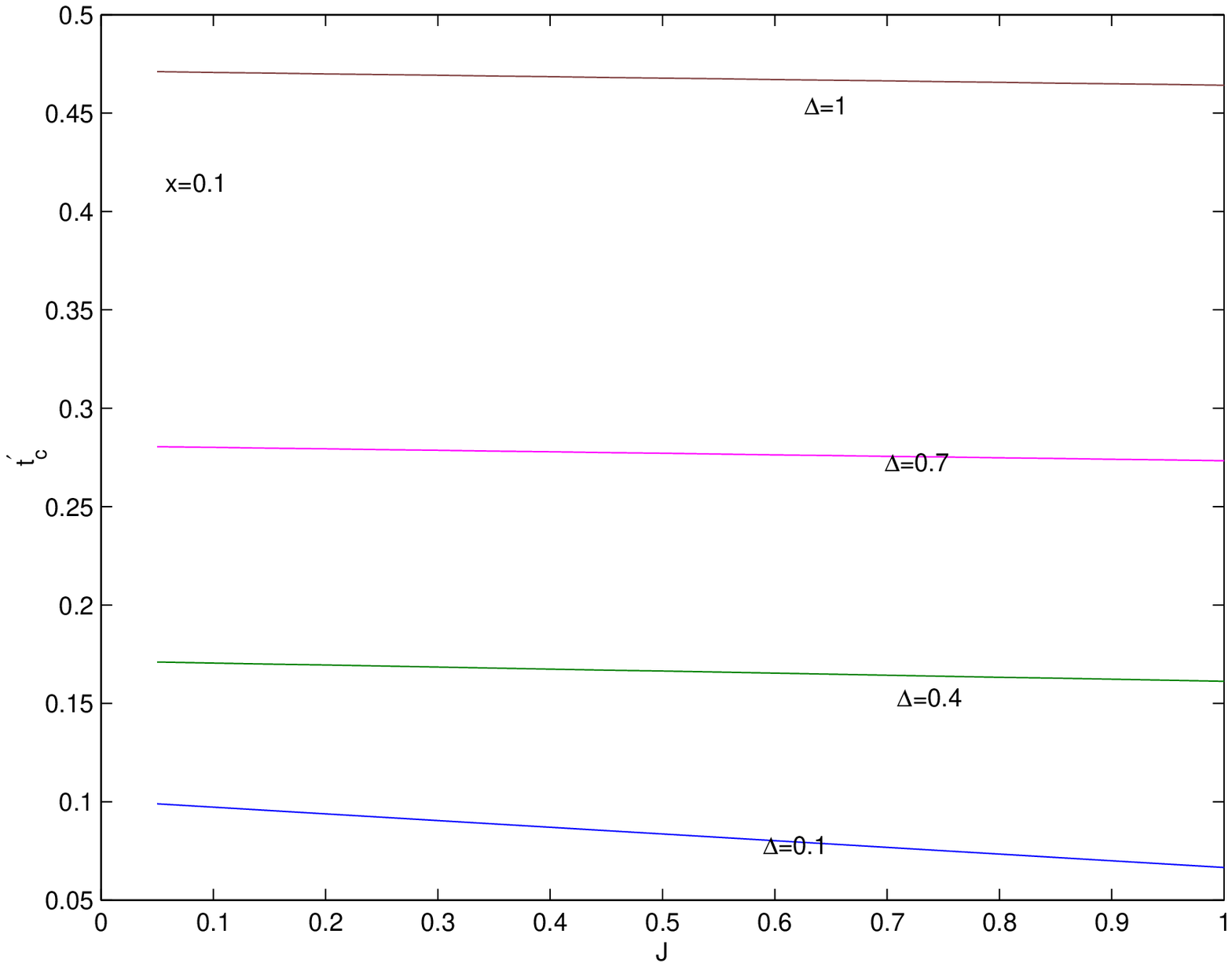}
\caption{The critical value of $t^{\prime}_c$ at $x=0.1$ and $t=1$
for different values of the deformed parameter $\Delta$.}
\label{phase1}
\end{center}
\end{figure}
\begin{figure}
\begin{center}
\includegraphics[width=0.9\columnwidth]{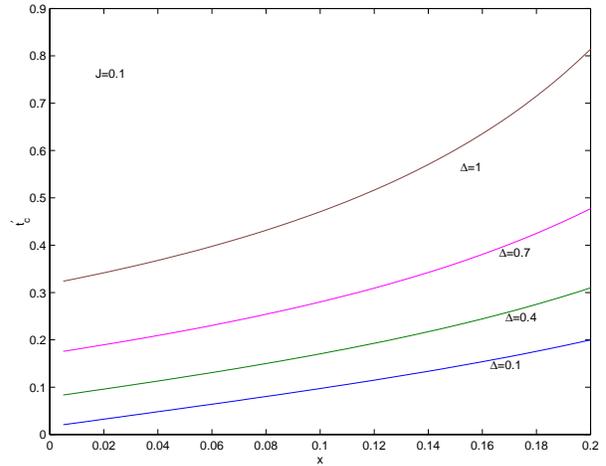}
\caption{The critical value of $t^{\prime}_c$ at $J=0.1$ and $t=1$
for different values of the deformed parameter $\Delta$.}
\label{phase2}
\end{center}
\end{figure}

To proceed, we have to use the expression of $\lambda_s$
(\ref{sp1}). For given $x>0$, $J/t\ll 1$, and $\Delta$, we found
that $\lambda_s$ will change its sign from $\lambda_s<0$ to
$\lambda_s>0$ by increasing the value of $t^{\prime}/t$, where
$x=1-n$ is the hole concentration. Although equation (\ref{sp1})
is derived under the assumption $\Delta \ll 1$, we shall
extrapolate it to $\Delta =1$ to get a rough estimate of the phase
boundary. The critical value $(t^{\prime}/t)_c$ is determined by
the equation $\lambda_s=0$. The solution of it gives the
dependence of $(t^{\prime}/t)_c$ on $x$ and $J/t$, and the results
are depicted in figures \ref{phase1} and \ref{phase2}. As
mentioned at the end of Sec. \ref{ttj}, the proper starting point
to treat the $1$D $t-t^{\prime}-J$ model is the two-band model
instead of the one-band model we employed in this paper for
$|t^{\prime}/t|>0.25$ and moderate values of $x$. Therefore, we
only consider the case with low doping concentrations in both
figures. Figure \ref{phase1} gives $(t^{\prime}/t)_c$ as a
function of $J/t$ at $x=0.1$ for $\Delta =0.1,0.4,0.7,1$. For
fixed $x$ and $\Delta$, $(t^{\prime}/t)_c$ is a decreasing
function of $J/t$. But in the range we considered it is, in fact,
insensitive to the value of $J/t$ at low doping concentration
except for small values of $\Delta$. On the other hand, for fixed
$J/t$ and $\Delta$, the hole concentration $x$ has dramatic
effects on $(t^{\prime}/t)_c$. This can be seen in figure
\ref{phase2}, which shows that $(t^{\prime}/t)_c$ is an increasing
function of $x$.

The above study indicates that the low-energy physics of the
deformed $1$D $t-t^{\prime}-J$ model is described by the LLs for
$t^{\prime}/t<(t^{\prime}/t)_c$, and it becomes a LE liquid as
$t^{\prime}/t>(t^{\prime}/t)_c$. In addition, figures \ref{phase1}
and \ref{phase2} suggest that $(t^{\prime}/t)_c>0$ even for
$\Delta =1$. In fact, by extrapolating equation (\ref{sp1}) into
$\Delta =1$, we found $(t^{\prime}/t)_c\approx 0.32$ as
$x\rightarrow 0^+$. (Note that the $1$D $t-J$ model falls into the
LL phase\cite{foot3}.) Accordingly, a positive $t^{\prime}$ favors
the opening of a spin gap and thus suppresses the AF ordering.
Moreover, the $J$ term plays a minor role in determining the phase
diagram at lightly doping as long as the system is far away from
the region of phase separation. This can be seen by expanding
equation (\ref{sp1}) according to the power of $x$. Then, the
dependence of $\lambda_s$ on $J/t$ starts from $O(x^2)$, whereas
its dependence on $t^{\prime}/t$ starts from $O(1)$.

\section{Spin-gap phase}
\label{sgp}

To further understand the properties of the ground state in the
spin-gap phase, we first examine the $2k_F$ CDW and SS
susceptibilities. The corresponding order parameters are defined
by
\begin{eqnarray}
 O_{CDW} &=& \psi^{\dagger}_{L\alpha}\psi_{R\alpha} \ , \nonumber
         \\
 O_{SS} &=& \frac{i}{2}~\epsilon_{\alpha \beta}\psi_{L\alpha}
        \psi_{R\beta} \ . \label{op1}
\end{eqnarray}
In terms of the bosonization formulas
\begin{equation}
 \psi_{L(R)\alpha}=\frac{1}{\sqrt{2\pi a_0}}~\eta_{\alpha}\exp{
                  \left\{\mp i\sqrt{4\pi}\phi_{L(R)\alpha}\right\}}
                  \ , \label{bos3}
\end{equation}
where $\eta_{\uparrow (\downarrow)}$ are real fermions which
satisfy $\eta_{\uparrow (\downarrow)}^2=1$, they can be bosonized
as
\begin{eqnarray}
 O_{CDW} &=& \frac{i\gamma}{\pi a_0}\exp{\left\{i\sqrt{2\pi}\Phi_c
         \right\}} \ , \nonumber \\
 O_{SS} &=& \frac{i\gamma}{2\pi a_0}\eta_{\uparrow}\eta_{\downarrow}
        \exp{\left\{i\sqrt{2\pi}\Theta_c\right\}} \ , \label{op2}
\end{eqnarray}
where $\Phi_c=(\phi_{\uparrow}+\phi_{\downarrow})/\sqrt{2}$,
$\Phi_s=(\phi_{\uparrow}-\phi_{\downarrow})/\sqrt{2}$, and
$\gamma=\langle \cos{(\sqrt{2\pi}\Phi_s)}\rangle$. Using equation
(\ref{op2}), we obtain the long distance behaviors of the $2k_F$
CDW and SS susceptibilities
\begin{eqnarray}
 \left\langle O_{CDW}(\tau ,x)O^{\dagger}_{CDW}(0,0)\right\rangle
 &\sim& \frac{1}{|z|^{K_c}} \ , \nonumber \\
 \left\langle O_{SS}(\tau ,x)O^{\dagger}_{SS}(0,0)\right\rangle
 &\sim& \frac{1}{|z|^{K_c^{-1}}} \ , \label{sus1}
\end{eqnarray}
where $z=v_c\tau +ix$. Equation (\ref{sus1}) indicates that the
dominant fluctuations are the SS for $K_c>1$ and the $2k_F$ CDW
for $K_c<1$. To determine which one occurs in the spin-gap phase,
we resort to equation (\ref{cp1}). In terms of $\lambda_s$,
$\lambda_c$ can be expressed by
\begin{equation}
 \lambda_c=-\frac{1}{4}~\lambda_s -\frac{4\Delta^2}{\pi}~
           \cos{\left(\frac{\pi}{2}x\right)}f(x) \ , \label{cp3}
\end{equation}
where $f(x)=t[1+\cos{(\pi
x)}]+4t^{\prime}\sin{\left(\frac{\pi}{2}x\right)}[1-\cos{(2\pi
x)}]$. Note that $f(x)>0$ for $0<x<1$ and $t^{\prime}>0$. The
spin-gap phase corresponds to $\lambda_s>0$, which results in
$\lambda_c<0$ from equation (\ref{cp3}). Plugging this into
equation (\ref{cp2}) gives $K_c>1$. Although both the SS and
$2k_F$ CDW correlations are enhanced in the spin-gap phase
compared with the free fermions, the SS one is dominant.

To seek the origin of the enhancement of the SS correlations in
the spin-gap phase, we notice that the structure of $H_s$
(\ref{hs1}) is identical to the continuum one for the spin-$1/2$
Heisenberg chain with the additional next-nearest-neighbor
exchange interaction $J^{\prime}$ (the $J-J^{\prime}$
model)\cite{Hal}. In that case, the low energy physics is
described by the LL as $J^{\prime}/J<(J^{\prime}/J)_c$, where
$(J^{\prime}/J)_c$ denotes the critical value of $J^{\prime}/J$,
because a negative value of $\lambda_s$ flows to zero under the RG
transformations. By increasing the value of $J^{\prime}$ such that
$J^{\prime}/J>(J^{\prime}/J)_c$ and thus $\lambda_s$ becomes
positive, then $\lambda_s(l\rightarrow \infty)\rightarrow \infty$
and the spin gap is opened\cite{Hal}. In the latter situation, the
opening of the spin gap is associated with the occurrence of the
spin-Peierls ordering, which breaks the symmetry of translation by
one site (the Z$_2$ symmetry) spontaneously. This comparison
implies that in the presence of a positive $t^{\prime}$, the spin
sector of the $t-t^{\prime}-J$ model has the tendency toward the
spin-Peierls ordering {\it without} including the $J^{\prime}$
term. (The spontaneous breaking of the Z$_2$ symmetry in the spin
background is restored by the hole motion in the present case. See
equation (\ref{dim2}).) On the other hand, the lowest-energy spin
excitations in the spin-gap phase are also the massive spinons,
which are the kink or anti-kink connecting two degenerate ground
states. This observation on the spin excitation spectrum provides
another evidence to support that the underlying spin background in
the spin-gap phase is the spin-Peierls state. Finally, we examine
the correlations of the dimerization operator:
\begin{equation}
 \epsilon (x)=\bm{S}_j\cdot \bm{S}_{j+1}-\bm{S}_{j+1}\cdot
             \bm{S}_{j+2} \ , \label{dim1}
\end{equation}
where $x=ja_0$. Then, for $|x|\rightarrow \infty$, we have
\begin{equation}
 \langle \epsilon (x)\epsilon (0)\rangle \sim \gamma^2
         \frac{\cos{(2k_Fx)}}{x^{K_c}} \ . \label{dim2}
\end{equation}
Since $2k_F=\pi (1-x)$, at lightly doping ($x\ll 1$), the cosine
term in equation (\ref{dim2}) locally mimics the alternation that
is characteristic of a dimerized spin chain (the $J-J^{\prime}$
model). The corresponding spin-Peierls ordering is weakened by a
power law due to the charge fluctuations.

To sum up, the effect of a positive $t^{\prime}$ is to enhance the
spin-Peierls ordering (and suppress the AF ordering) such that the
whole system behaves like a doped spin-Peierls state. The doped
holes in the spin-Peierls state are inclined to form the local
hole pairs due to the energetic consideration. Furthermore, the
calculation of the SS correlator indicates that the other effect
of a positive $t^{\prime}$ is to enhance the Josephson tunneling
between these local hole pairs, which results in the phase
coherence between them\cite{MXA}.

\section{Conclusions and Discussions}
\label{last}

We obtain the phase diagram of the $1$D $t-t^{\prime}-J$ model by
studying the deformed version of it. We found that the effects of
the positive $t^{\prime}$ at lightly doping are (i) to open the
spin gap by enhancing the spin-Peierls ordering (and suppressing
the AF ordering), and (ii) to make the hole-pair propagation
become coherent and thus enhance the SS correlations. In other
words, a positive $t^{\prime}$ provides a mechanism for the
occurrence of the $1$D superconductors in the {\it one-band}
system with purely repulsive force.

Our findings about the effects of a positive $t^{\prime}$ are
reminiscent of the role of the next-nearest-neighbor exchange
interaction, $J^{\prime}$, played on the spin-$1/2$ Heisenberg
chain. The existence of a spin-gap phase in the $1$D
$t-J-J^{\prime}$ model at low doping concentration was indeed
predicted, as long as $J^{\prime}/J>(J^{\prime}/J)_c\approx 0.24$
and $J/t<(J/t)_p$, where $(J/t)_p=O(1)$ denotes the boundary of
phase separation\cite{OLR,SL}. The underlying reason is that a
spin gap already exists at half-filling for
$J^{\prime}/J>(J^{\prime}/J)_c$ and it is stable against the
lightly hole doping. As a matter of fact, the opening of the spin
gap in both the $t-t^{\prime}-J$ and $t-J-J^{\prime}$ models
results from the same operator. From the viewpoint of the Hubbard
model, i.e. the $t-t^{\prime}-U$ model\cite{foot1}, a $J^{\prime}$
term in the low-energy effective Hamiltonian will be generated
from the $t^{\prime}$ term in the large-$U$ limit, where $U>0$
denotes the on-site interactions between electrons. However, the
value of $J^{\prime}$ generated by this way is given by
$J^{\prime}/J\sim (t^{\prime}/t)^2$. According to our results
(figure \ref{phase2}), the induced $J^{\prime}$ is still in the
region $J^{\prime}/J<(J^{\prime}/J)_c$ at lightly doping when
$t^{\prime}/t>(t^{\prime}/t)_c$. Therefore, it is less possible
that the spin-gap phase we found arises from the $J^{\prime}$ term
induced by a positive $t^{\prime}$ though the spin-gap phases in
both models belong to the same universality class.

It should be mentioned that our results about the opening of the
spin gap in the deformed $t-t^{\prime}-J$ chain are also valid at
half-filling, i.e. $n=1$. In that case, the Umklapp processes
ignored in our analysis become relevant and thus a charge gap is
opened, which corresponds to $K_c=0$. Because the Umklapp
processes are composed of the spin-singlet operators, they do not
affect the spin sector, especially equation (\ref{sp1}).
Therefore, there still exists $(t^{\prime}/t)_c>0$ at half-filling
such that there is an algebraic long range AF order with gapless
spin excitations for $t^{\prime}/t<(t^{\prime}/t)_c$, whereas a
spin liquid phase with the long-ranged spin-Peierls order emerges
for $t^{\prime}/t>(t^{\prime}/t)_c$. (Note that at half-filling,
equation (\ref{dim2}) gives $\langle \epsilon (x)\rangle \sim
(-1)^x\gamma$.) In this sense, the deformed $t-t^{\prime}-J$ model
at half-filling behaves like the $J-J^{\prime}$ model instead of
the Heisenberg chain (or the $t-t^{\prime}-J$ chain at
half-filling), which is known to be gapless in its spin sector. As
mentioned in the previous paragraph, an effective $J^{\prime}$
term will be generated from the $t^{\prime}$ term in the strong
coupling regime of the $t-t^{\prime}-U$ model, which also contains
both the Luttinger liquid and spin-gap phases. Compared with our
results, it seems to suggest that in the case of half-filling the
deformed $t-t^{\prime}-J$ model may belong to the same
universality class of the $t-t^{\prime}-U$ model in the strong
coupling regime, in stead of the $t-t^{\prime}-J$ model. Whether
this discrepancy at half-filling is a generical feature or not
needs more work to clarify it.

The application of our analysis on the deformed $t-J$ type model
to the usual one ($\Delta =1$) is based on the assumption that the
deformed model with $\Delta \ll 1$ is adiabatically connected to
the one with $\Delta =1$. That is, there is no phase transition by
increasing the deformation parameter $\Delta$ from $\Delta \ll 1$
to $\Delta =1$. This assumption has been examined in reference
\cite{CW} where the phase diagram of the $t-J_z$ model was
studied. The phase diagram obtained by the deformed model is
identical to that predicted by the numerical methods. This
supports the use of the technique of the deformed Hubbard operator
to study the $t-J$ type models. However, a recent numerical work
on the $t-t^{\prime}-J$ chain indicated that even moderate values
of $t^{\prime}$ results in the breakdown of the LLs\cite{EO}. In
addition, the low-doping phase shows similarities with the doped
two-leg $t-J$ ladders where the Fermi surface takes the form of a
hole pocket and the quantum numbers carried by the elementary
excitations are the same as those of the doped holes. From the
point of view of the renormalization group, the instability of the
LLs found in reference \cite{EO} should result from a relevant
operator generated by the $t^{\prime}$ term, which mediates an
attraction between the charge and spin sectors. For example, the
spin-charge recombination in the two-leg ladders arises from the
superexchange interaction along the rung. The operators which mix
the charge and spin sectors may occur in the six-fermion
interactions in the continuum theory. But these are irrelevant
operators in the weak coupling regime, i.e $\Delta \ll 1$.
Nevertheless, we cannot exclude the possibility that one of the
six-fermion interactions mixing the charge and spin sectors
becomes relevant in the strong regime, i.e. $\Delta =1$. This
problem is intimately related to the previous one: whether a phase
transition exists between $\Delta \ll 1$ and $\Delta =1$ or not.
This issue deserves further study. In addition, to determine the
exact value of $(t^{\prime}/t)_c$ at $\Delta =1$, further
numerical work is warranted.

{\small We would like to thank the discussions with Y.C. Chen,
M.F. Yang, and T.K. Lee. Y.L. L. is grateful to Global
fiberoptics, Inc. for financial support. The work of Y.-W. L. is
supported by the National Science Council of Taiwan under grant
NSC 91-2112-M-029-012.}

\appendix
\section{Derivation of the continuum Hamiltonian}
\label{ham}
\renewcommand{\theequation}{\Alph{section}\arabic{equation}}
\setcounter{equation}{0}

In this appendix, we provide the details for the derivation of the
continuum Hamiltonian $H_{\psi}$ from the lattice model given by
equations (\ref{ht}) and (\ref{hj}). Equations (\ref{cp1}) and
(\ref{sp1}) result from this procedure automatically.

Inserting equation (\ref{hop2}) into equation (\ref{ht}) gives
\begin{eqnarray*}
 {\rm H}_h={\rm H}_0+{\rm H}_1+{\rm H}_2+{\rm H}_3+{\rm H}_4 \ ,
\end{eqnarray*}
where
\begin{eqnarray}
 {\rm H}_0 &=& -t\left(1-\Delta \frac{n}{2}\right)^2\sum_{j,\alpha}
           \left(c_{j+1\alpha}^{\dagger}c_{j\alpha}+{\rm H.c.}\right)
           \nonumber \\
           & & -t^{\prime}\left(1-\Delta \frac{n}{2}\right)^2
           \sum_{j,\alpha}\left(c_{j+2\alpha}^{\dagger}c_{j\alpha}
           +{\rm H.c.}\right) \ , \label{ht0}
\end{eqnarray}
and
\begin{eqnarray}
 {\rm H}_1 &=& t\Delta \left(1-\Delta \frac{n}{2}\right)\sum_{j,\alpha}
           \left[c^{\dagger}_{j+1\alpha}c_{j\alpha}(:n_{j-\alpha}:
           +:n_{j+1-\alpha}:)\right. \nonumber \\
           & & \left.+{\rm H.c.}\right] \ , \label{ht1} \\
 {\rm H}_2 &=& -t\Delta^2\sum_{j,\alpha} \! \! \left(c^{\dagger}_{j+1\alpha}
           c_{j\alpha} \! \! :n_{j+1-\alpha}::n_{j-\alpha}: \! \!
           +{\rm H.c.}\right) \ , \label{ht2} \\
 {\rm H}_3 &=& t^{\prime}\Delta \left(1-\Delta \frac{n}{2}\right)
           \sum_{j,\alpha}\left[c^{\dagger}_{j+2\alpha}c_{j\alpha}
           (:n_{j-\alpha}:+:n_{j+2-\alpha}:)\right. \nonumber \\
           & & \left.+{\rm H.c.}\right] \ , \label{ht3} \\
 {\rm H}_4 &=& -t^{\prime}\Delta^2\sum_{j,\alpha} \! \! \left(
           c^{\dagger}_{j+2\alpha}c_{j\alpha} \! \! :n_{j+2-\alpha}:
           :n_{j-\alpha}: \! \! +{\rm H.c.}\right) \ . \label{ht4}
\end{eqnarray}
In the above, $:n_{j\alpha}:=n_{j\alpha}-\frac{n}{2}$ and $n$ is
the electron density. In the following, we shall restrict to the
case away from half-filling.

We first consider ${\mathrm H}_1$. Using equation (\ref{eop1}),
one may find the following operator product expansion (OPE) for
$\alpha =\uparrow ,\downarrow$
\begin{eqnarray}
 c_{j+1\alpha}^{\dagger}c_{j\alpha}/a_0 &\approx& \frac{\sin{(k_Fa_0)}}
   {\pi a_0}+e^{ik_Fa_0}J_{L\alpha}+e^{-ik_Fa_0}J_{R\alpha} \nonumber
   \\
   & & +\left(e^{2ik_Fx}e^{ik_Fa_0}\psi^{\dagger}_{L\alpha}\psi_{R\alpha}
   +{\mathrm H.c.}\right) . \label{ope5}
\end{eqnarray}
Furthermore, in terms of the $\psi$-fermions, the density operator
for spin $\alpha$, $:n_{j\alpha}:$, can be written as
\begin{equation}
 :n_{j\alpha}:\approx a_0\left[J_{L\alpha}+J_{R\alpha}+\left(e^{2ik_Fx}
              \psi^{\dagger}_{L\alpha}\psi_{R\alpha}+{\mathrm H.c.}
              \right)\right] \ , \label{den1}
\end{equation}
where
$J_{L(R)\alpha}=:\psi^{\dagger}_{L(R)\alpha}\psi_{L(R)\alpha}:$.
With the help of equations (\ref{ope5}) and (\ref{den1}) and
neglecting the constant term, ${\mathrm H}_1$ becomes
\begin{eqnarray*}
 {\mathrm H}_1 &=& 8ta_0\Delta \left(1-\Delta \frac{n}{2}\right)\cos{(k_Fa_0)}
           \! \! \int \! \! dx\left(J_{L\uparrow}J_{L\downarrow}
           +J_{R\uparrow}J_{R\downarrow}\right. \\
           & & \left.+J_LJ_R-:\psi_{L\alpha}^{\dagger}\psi_{L\beta}
           \psi_{R\beta}^{\dagger}\psi_{R\alpha}:\right) \ .
\end{eqnarray*}
In terms of the identity
\begin{equation}
 J_L^zJ_L^z=\frac{1}{3}\bm{J}_L\cdot \bm{J}_L \ , \label{id7}
\end{equation}
we obtain
\begin{equation}
 J_{L\uparrow}J_{L\downarrow}=\frac{1}{4}J_LJ_L-J_L^zJ_L^z
   =\frac{1}{4}J_LJ_L-\frac{1}{3}\bm{J}_L\cdot \bm{J}_L \ .
   \label{id1}
\end{equation}
In addition, we have
\begin{equation}
 :\psi_{L\alpha}^{\dagger}\psi_{L\beta}\psi_{R\beta}^{\dagger}
 \psi_{R\alpha}:=\frac{1}{2}J_LJ_R+2\bm{J}_L\cdot \bm{J}_R \ .
 \label{id2}
\end{equation}
Inserting equations (\ref{id1}) and (\ref{id2}) into ${\rm H}_1$
gives
\begin{eqnarray}
 & & {\mathrm H}_1=2ta_0\Delta \left(1-\Delta \frac{n}{2}\right)
     \cos{\left(\frac{\pi}{2}n\right)} \! \! \int \! \! dx \{(
     : \! \! J_LJ_L \! \! :+: \! \! J_RJ_R \! \! : \nonumber \\
 & & +2J_LJ_R)-\frac{4}{3}(: \! \! \bm{J}_L\cdot \bm{J}_L \! \! :
     +: \! \! \bm{J}_R\cdot \bm{J}_R \! \! :+6\bm{J}_L\cdot
     \bm{J}_R)\} \ . \label{htd1}
\end{eqnarray}
The same procedure is applied to ${\rm H}_3$ except that $t$ and
$a_0$ are replaced by $t^{\prime}$ and $2a_0$, respectively.
Accordingly, we have
\begin{eqnarray}
 & & {\mathrm H}_3=4t^{\prime}a_0\Delta \left(1-\Delta \frac{n}{2}
     \right)\cos{(\pi n)} \! \! \int \! \! dx \{(: \! \! J_LJ_L \! \! :
     +: \! \! J_RJ_R \! \! : \nonumber \\
 & & +2J_LJ_R)-\frac{4}{3}(: \! \! \bm{J}_L\cdot \bm{J}_L \! \! :
     +: \! \! \bm{J}_R\cdot \bm{J}_R \! \! :+6\bm{J}_L\cdot \bm{J}_R)
     \} \ . \label{htd3}
\end{eqnarray}

Next we investigate ${\mathrm H}_2$. To proceed, we need the
following OPE's:
\begin{eqnarray}
 J_{L\alpha}(z)\psi_{L\beta}(0) &\sim& -\frac{\delta_{\alpha \beta}}
            {2\pi z}\psi_{L\alpha}(0) \ , \label{ope1} \\
 J_{L\alpha}(z)\psi_{L\beta}^{\dagger}(0) &\sim& \frac{
            \delta_{\alpha \beta}}{2\pi z}\psi_{L\alpha}^{\dagger}(0)
            \ . \nonumber
\end{eqnarray}
Then, using equations (\ref{den1}) and (\ref{ope1}), we obtain for
$\alpha =\uparrow ,\downarrow$
\begin{eqnarray}
 & & :n_{j+1\alpha}::n_{j\alpha}:/a_0^2 \nonumber \\
 & & \approx -\frac{2\sin{(k_Fa_0)}}{\pi a_0}\left\{\frac{\sin{(k_Fa_0)}}
     {2\pi a_0}+\cos{(k_Fa_0)}(J_{L\alpha}+J_{R\alpha})\right. \nonumber
     \\
 & & \left.+\left(e^{2ik_Fx}e^{ik_Fa_0}\psi_{L\alpha}^{\dagger}
     \psi_{R\alpha}+{\rm H.c.}\right)\right\}+:J_{L\alpha}J_{L\alpha}:
     \nonumber \\
 & & +:J_{R\alpha}J_{R\alpha}:+4\sin^2{(k_Fa_0)}J_{L\alpha}J_{R\alpha}
     \ . \label{ope2}
\end{eqnarray}
In equation (\ref{ope2}), the $2nk_F$ terms with $n>1$ are
neglected. By using equations (\ref{ope5}) and (\ref{ope2}) and
neglecting a constant term, ${\mathrm H}_2$ becomes
\begin{eqnarray*}
 & & {\mathrm H}_2 \! = \! \frac{8ta_0\Delta^2 \! \sin{(k_Fa_0)}}{\pi} \! \!
     \int \! \! dx \! \left\{\cos^2{ \! (k_Fa_0)}( \! J_{L\uparrow}
     J_{L\downarrow} \! \! + \! \! J_{R\uparrow}J_{R\downarrow}\right. \\
 & & \left.+ \! J_LJ_R) \! - \! \! \frac{1}{4}(: \! \! J_{L\alpha}J_{L\alpha}
     \! \! : \! + \! : \! \! J_{R\alpha}J_{R\alpha} \! \! :)- \! \! : \! \!
     \psi_{L\alpha}^{\dagger}\psi_{L\beta}\psi_{R\beta}^{\dagger}
     \psi_{R\alpha} \! \! :\right\} .
\end{eqnarray*}
With the help of equations (\ref{id1}), (\ref{id2}), and the
identity
\begin{eqnarray}
 & & :J_{L\alpha}J_{L\alpha}:+:J_{R\alpha}J_{R\alpha}: \label{id6} \\
 & & =\frac{1}{2}(: \! \! J_LJ_L \! \! :+: \! \! J_RJ_R \! \! :)
     +\frac{2}{3}\left(: \! \! \bm{J}_L\cdot\bm{J}_L \! \! :+: \! \!
     \bm{J}_R\cdot \bm{J}_R \! \! :\right) \ , \nonumber
\end{eqnarray}
we arrive at
\begin{eqnarray}
 {\mathrm H}_2 &=& \frac{ta_0\Delta^2\sin{(\pi n/2)}\cos{(\pi n)}}{\pi}
               \! \! \int \! \! dx (:J_LJ_L:+:J_RJ_R: \nonumber \\
               & & +4J_LJ_R)-\frac{4ta_0\Delta^2\sin{(\pi n/2)}}{3\pi}
               \! \! \int \! \! dx \{[2+\cos{(\pi n)}] \nonumber \\
               & & \times (:\bm{J}_L\cdot \bm{J}_L:+:\bm{J}_R\cdot
               \bm{J}_R:)+12\bm{J}_L\cdot \bm{J}_R\} \ . \label{htd2}
\end{eqnarray}
With the same procedure, ${\mathrm H}_4$ becomes
\begin{eqnarray}
 {\mathrm H}_4 &=& \frac{2t^{\prime}a_0\Delta^2\sin{(\pi n)}\cos{(2\pi n)}}
               {\pi} \! \! \int \! \! dx(:J_LJ_L:+:J_RJ_R: \nonumber \\
               & & +4J_LJ_R)-\frac{8t^{\prime}a_0\Delta^2\sin{(\pi n)}}{3\pi}
               \! \! \int \! \! dx\{[2+\cos{(2\pi n)}] \nonumber \\
               & & \times (:\bm{J}_L\cdot \bm{J}_L:+:\bm{J}_R\cdot \bm{J}_R:)
               +12\bm{J}_L\cdot \bm{J}_R\} \ . \label{htd4}
\end{eqnarray}

Finally, we turn into ${\mathrm H}_J$. First, we have
\begin{eqnarray}
 & & :n_{j+1}::n_j:/a_0^2 \nonumber \\
 & & \approx -\frac{2\sin{(k_Fa_0)}}{\pi a_0}\left[\frac{\sin{(k_Fa_0)}}
     {\pi a_0}+\cos{(k_Fa_0)}(J_L+J_R)\right] \nonumber \\
 & & +:J_LJ_L:+:J_RJ_R:+2J_LJ_R-2\cos{(2k_Fa_0)} \nonumber \\
 & & \times :\psi_{L\alpha}^{\dagger}\psi_{L\beta}\psi_{R\beta}^{\dagger}
     \psi_{R\alpha}: \nonumber \\
 & & \approx -\frac{2\sin{(k_Fa_0)}}{\pi a_0}\left[\frac{\sin{(k_Fa_0)}}
     {\pi a_0}+\cos{(k_Fa_0)}(J_L+J_R)\right] \nonumber \\
 & & +:J_LJ_L:+:J_RJ_R:+[2-\cos{(2k_Fa_0)}]J_LJ_R \nonumber \\
 & & -4\cos{(2k_Fa_0)}\bm{J}_L\cdot \bm{J}_R \ , \label{ope3}
\end{eqnarray}
Next, in terms of the $\psi$-fermions, the spin operator can be
written as
\begin{equation}
 \bm{S}_j\approx a_0\left[\bm{J}_L+\bm{J}_R+\left(e^{2ik_Fx}\bm{m}
         +{\mathrm H.c.}\right)\right] \ , \label{spin1}
\end{equation}
where
$\bm{m}=\frac{1}{2}\psi_{L\alpha}^{\dagger}(\bm{\sigma})_{\alpha
\beta}\psi_{R\beta}$. Using the OPE's
\begin{eqnarray}
 \bm{m}(x+a_0)\cdot \bm{m}^{\dagger}(x) &=& \frac{3}{8\pi a_0}\left[
       \frac{1}{\pi a_0}+i(J_L+J_R)\right] \nonumber \\
       & & +\frac{1}{2}\bm{J}_L\cdot \bm{J}_R-\frac{3}{8}J_LJ_R \ ,
       \label{id3} \\
 \bm{m}^{\dagger}(x+a_0)\cdot \bm{m}(x) &=& \frac{3}{8\pi a_0}\left[
       \frac{1}{\pi a_0}-i(J_L+J_R)\right] \nonumber \\
       & & +\frac{1}{2}\bm{J}_L\cdot \bm{J}_R-\frac{3}{8}J_LJ_R \ ,
       \nonumber
\end{eqnarray}
we obtain
\begin{eqnarray}
 & & \bm{S}_{j+1}\cdot \bm{S}_j/a_0^2 \nonumber \\
 & & \approx -\frac{3}{4\pi^2a^2_0}+:\bm{J}_L\cdot \bm{J}_L:
     +:\bm{J}_R\cdot \bm{J}_R:+2\bm{J}_L\cdot \bm{J}_R \nonumber
     \\
 & & +e^{2ik_Fa_0}\bm{m}(x+a_0) \! \cdot \! \bm{m}^{\dagger}(x)+
     e^{-2ik_Fa_0}\bm{m}^{\dagger}(x+a_0) \! \cdot \! \bm{m}(x)
     \nonumber \\
 & & \approx -\frac{3\sin{(k_Fa_0)}}{2\pi a_0}\left[\frac{
     \sin{(k_Fa_0)}}{\pi a_0}+\cos{(k_Fa_0)}(J_L+J_R)\right]
     \nonumber \\
 & & +:\bm{J}_L\cdot \bm{J}_L:+:\bm{J}_R\cdot \bm{J}_R:
     +[2+\cos{(2k_Fa_0)}]\bm{J}_L\cdot \bm{J}_R \nonumber \\
 & & -\frac{3}{4}\cos{(2k_Fa_0)}J_LJ_R \ . \label{ope4}
\end{eqnarray}
Again, the $2nk_F$ terms with $n>1$ in equations (\ref{ope3}) and
(\ref{ope4}) are neglected. With the help of equations
(\ref{ope3}) and (\ref{ope4}), we get
\begin{eqnarray}
 & & {\mathrm H}_J=Ja_0 \! \! \int \! \! dx\left\{: \! \! \bm{J}_L
     \cdot \bm{J}_L \! \! :+: \! \! \bm{J}_R\cdot \bm{J}_R \! \! :
     +4\cos^2{ \! \! \left(\frac{\pi}{2}n\right)}\right. \nonumber
     \\
 & & \left. \times \bm{J}_L\cdot \bm{J}_R\right\}-\frac{Ja_0}{4}
     \! \! \int \! \! dx\left\{: \! \! J_LJ_L \! \! :+: \! \! J_RJ_R
     \! \! :+4\cos^2{ \! \! \left(\frac{\pi}{2}n\right)}\right.
     \nonumber \\
 & & \left. \times J_LJ_R\right\} \ . \label{hjd}
\end{eqnarray}
By collecting equations (\ref{htd1}), (\ref{htd3}), (\ref{htd2}),
(\ref{htd4}), and (\ref{hjd}), we obtain the continuum Hamiltonian
$H_{\psi}$ with the expressions (\ref{cp1}) and (\ref{sp1}).



\begin{thebibliography}{99}
\bibitem{foot3} We also studied the spin-anisotropic $t-J$ model,
                 i.e. the $t-J_{\perp}-J_z$ model, by using the approach
                 of the deformed Hubbard operator. The continuum
                 Hamiltonian we obtained is similar to $H_{\psi}$,
                 except that the $\lambda_s$ term in equation (\ref{hs1})
                 becomes $\frac{\lambda_{\perp}}{2}(J_L^+J_R^-
                 +J_L^-J_R^+)+\lambda_3J_L^3J_R^3$. Up to four-fermion
                 interactions, we do not find the couplings between the
                 charge and spin sectors. When $J_{\perp}\neq J_z$
                 ($\lambda_{\perp}\neq \lambda_3$), there are two phases:
                 the Ising-N\'eel phase and the spin-Peierls phase. Both
                 have a finite spin gap due to the breaking of the spin
                 SU($2$) symmetry. We do not find the duality symmetry
                 claimed in reference \cite{CW}.
\end{thebibliography}
\end{document}